\documentclass[twocolumn,aps,showpacs,prb]{revtex4-1}

\usepackage{amsmath,amssymb,graphics,epsfig,epstopdf,color,multirow,array,verbatim,ulem,braket,tabularx}
\usepackage[colorlinks,linkcolor=blue,citecolor=blue,urlcolor=blue]{hyperref}
\usepackage{color}
\usepackage{graphicx}
\usepackage{epstopdf}
\usepackage{amsmath}
\usepackage{multirow}
\usepackage{ulem}

\begin{document}

\title{Superconductivity in monolayer Ba$_2$N electride: a first-principles study}

\author{Xiao-Le Qiu}
\author{Jian-Feng Zhang}
\author{Huan-Cheng Yang}
\author{Zhong-Yi Lu}\email{zlu@ruc.edu.cn}
\author{Kai Liu}\email{kliu@ruc.edu.cn}

\affiliation{Department of Physics and Beijing Key Laboratory of Opto-electronic Functional Materials $\&$ Micro-nano Devices, Renmin University of China, Beijing 100872, China}

\date{\today}

\begin{abstract}

The exploration of superconductivity in low-dimensional materials has attracted intensive attention for decades. Based on first-principles electronic structure calculations, we have systematically investigated the electronic and superconducting properties of the two-dimensional electride Ba$_2$N in the monolayer limit. Our results show that monolayer Ba$_2$N has a low work function of 3.0 eV and a predicted superconducting transition temperature ($T_c$) of 3.4 K. The superconductivity can be further improved with the tensile strain, which results from the increase of density of states at the Fermi level as well as the enhanced coupling between inner-layer electrons and phonons. Remarkably, at the 4$\%$ tensile strain, the acoustic branches have noticeable softening at the K point of Brillouin zone and the superconducting $T_c$ can reach 10.8 K.
The effect of lattice strain on the electron transfer from the superficial region to the inner-layer region of monolayer Ba$_2$N may also apply to other electride materials and influence their physical properties.

\end{abstract}


\maketitle

\section{INTRODUCTION}

The superconductivity in two-dimensional (2D) materials is a fascinating subject in condensed matter physics and has been extensively explored.
To achieve 2D superconductivity, many experimental studies have been carried out, for example, constructing interfacial superconductors~\cite{N. Reyren, J. Biscaras1, J. Biscaras2, J. Biscaras3, A. Gozar1, A. Gozar2,Z. Chen, Y.-Q. Sun, Q.-Y. Wang}, growing atomic metal layers~\cite{Y. Guo, T. Zhang, T. Uchihashi, M. Yamada}, reducing material dimensions~\cite{S. Foner, T. Yokoya, D. J. Rahn, X.-X. Xi, C. Sergio, D. Geng, Y. Yang}, and doping charges with ionic liquid gating or atom intercalation~\cite{J. T. Ye, Y. Saito2, J. T. Ye2, D. Costanzo, S. Jo}.
Due to the particular structure, the layered materials have unique advantage in dimension reduction. Through mechanical exfoliation~\cite{M. Yi}, liquid exfoliation~\cite{V. Nicolosi}, chemical vapor deposition (CVD)~\cite{D. Geng}, and molecular beam epitaxy (MBE)~\cite{D. Fu} methods, it is easy to reduce the dimension of layered materials from three-dimensional (3D) bulk crystals to 2D atomically thin films or monolayers. Meanwhile, the superconductivity of the layered materials may change along with the dimension reduction. Experimentally, the transition-metal dichalcogenide (TMD) 2H-NbSe$_2$ was mechanically exfoliated into monolayer~\cite{X.-X. Xi}, and at same time the superconducting $T_c$  drops to 3 K~\cite{X.-X. Xi, C. Sergio}, in comparison with 7 K in the bulk crystal~\cite{S. Foner, T. Yokoya, D. J. Rahn}. Similarly, monolayer 2H-TaS$_2$ was prepared via mechanical exfoliation method~\cite{Y.-F. Cao}. In contrast to 2H-NbSe$_2$, the $T_c$ of 2H-TaS$_2$ increases from 0.8 K in the bulk to 3.4 K in the monolayer~\cite{C. Sergio, Y. Yang}. Moreover, the ultrathin films of Mo$_2$C prepared by CVD method also show the dimension effect with the $T_c$ ranging from 3.5 K at 40 nm to 2.6 K at 5 nm~\cite{D. Geng}.
On the other hand, many theoretical proposals have been made to improve the superconducting temperatures of 2D superconductors. For example, the superconducting $T_c$ of freestanding borophene is predicted to be $\sim$20 K~\cite{E. S. Penev, R. C. Xiao} and may increase to 34.8 K via hole doping or 27.4 K by applying tensile strain~\cite{R. C. Xiao}. The calculated $T_c$ of monolayer MgB$_2$ is also enhanced from 20 K to 53 K with the tensile strain~\cite{J. Bekaert}, while the predicted $T_c$ of hydrogenated MgB$_2$ may reach 67 K and further boost to $>$ 100 K by biaxial tensile strain~\cite{J. Bekaert2}. Moreover, several 2D-MXene materials, such as W$_2$C, Sc$_2$C, Mo$_2$N, and Ta$_2$N, have also been predicted to be superconducting down to the monolayer forms, among which monolayer Mo2N has the highest $T_c$ of ~16 K~\cite{mo2n}. The rich physical phenomena concealed in the 2D superconductors still wait for further investigation.

Among various 2D superconductors, the 2D electride is an interesting material family, in which the extra electrons are confined in the interlayer region  acting as anions and the system exhibits inherent metallicity and high electron concentrations~\cite{K. Lee, X. Zeng, S. W. Kim}.
As a typical 2D electride, Ca$_2$N was synthesized in 2013~\cite{K. Lee, J. S. Oh} and can be exfoliated into 2D flakes via the liquid exfoliation method~\cite{D. L. Druffel}. Zeng $et$ $al.$ studied the electron-phonon interaction of Ca$_2$N monolayer theoretically and suggested that it is a Bardeen-Cooper-Schrieffer (BCS) superconductor with a superconducting $T_c$ of 4.7 K~\cite{X. Zeng}. Besides Ca$_2$N, some other 2D electride superconductors Y$_2$C and MgONa have also been proposed with the respective $T_c$'s of 0.9 K and 3.4 K~\cite{Y. Ge}. However, very few 2D electrides have been reported to be superconductors so far, and the predicted superconducting transition temperatures of them are relatively low. Therefore, it is necessary to find more 2D electride superconductors and endeavor to increase their superconducting temperature. Being isostructural to Ca$_2$N, Ba$_2$N is also a 2D electride~\cite{A. Walsh, T. Tada, S.-Y. Liu} that has been synthesized experimentally~\cite{O. Reckeweg}. Owing to its layered structure, Ba$_2$N may also have a potential to reduce to 2D flakes or monolayer via the liquid exfoliation method as Ca$_2$N~\cite{D. L. Druffel}. Therefore, we wonder whether monolayer Ba$_2$N is superconducting as well and whether the strain can affect the distribution of anionic electrons and its superconductivity. Moreover, due to the same sandwich-like structures of Ba$_2$N with 1T-TMDs and 2-1 MXenes, it is also interesting to make a comparation of their electronic and phonon properties, which may help gain an in-depth understanding of the related physical mechanism.

In this work, we have systematically studied the electronic and superconducting properties of the monolayer form of Ba$_2$N by using first-principles electronic structure calculations. We find that monolayer Ba$_2$N has a low work function of 3.0 eV due to the extra electrons on the surface and it is also a superconductor with the predicted $T_c$ of 3.4 K. With the application of tensile strain, the acoustic phonon branches soften remarkably around the K point, accompanied with a strong electron-phonon coupling. The superconducting $T_c$ can reach 10.8 K at the 4$\%$ tensile strain, while the lattice instability occurs when the strain exceeds 5$\%$.

\section{COMPUTATIONAL DETAILS}

The structural and electronic properties of monolayer Ba$_2$N were studied based on the density functional theory (DFT) calculations with the projector augmented wave (PAW) method~\cite{Blochl1994, Kresse1999} as implemented in the VASP package~\cite{vasp3, vasp1, vasp2}. The generalized gradient approximation (GGA) of the Perdew-Burke-Ernzerhof (PBE) type was adopted for the exchange-correlation functional~\cite{Perdew1996}. The energy cutoff of the plane wave basis was set to 520 eV. The Fermi surface was broadened by the Gaussian smearing method with a width of 0.05 eV. The convergence tolerances of force and energy were set to 0.01 eV/{\AA} and 10$^{-5}$ eV, respectively. To avoid the interactions between Ba$_2$N and its images, a vacuum layer larger than 15 {\AA} was added along the $z$ direction. A 12$\times$12$\times$1 Monkhorst-Pack {\bf k}-point mesh was used to sample the 2D Brillouin zone (BZ).

To investigate the phonon spectra and electron-phonon coupling (EPC), the density functional perturbation theory (DFPT)~\cite{S. Baroni} calculations were performed with the Quantum ESPRESSO (QE) package~\cite{QE}. The kinetic energy cutoffs of 60 and 600 Ry were chosen for the wavefunctions and the charge densities, respectively. The Fermi surface was also broadened by the Gaussian smearing method with a width of 0.05 eV (0.003675 Ry). The EPC constant was calculated by using a 6$\times$6$\times$1 {\bf q}-point mesh and a dense 72$\times$72$\times$1 {\bf k}-point mesh. The superconducting transition temperature $T_c$ was calculated with the McMillan-Allen-Dynes formula~\cite{McMillan1}:
\begin{equation}
T_c=\frac{\omega_{log}}{1.2}\text{exp}[\frac{-1.04(1+\lambda)}{\lambda(1-0.62\mu^*)-\mu^*}],
\end{equation}
where $\mu^*$ is the effective screened Coulomb repulsion constant which was set to an empirical value of 0.1~\cite{miu1, miu2}. The total EPC constant $\lambda$ can be obtained either by summing the EPC constant $\lambda_{{\bf q}\nu}$ for all phonon modes in the whole BZ or by integrating the Eliashberg spectral function $\alpha^2F(\omega)$ as follows~\cite{Eliashberg}
\begin{equation}
\lambda=\sum_{{\bf q}\nu}\lambda_{{\bf q}\nu}=2\int{\frac{\alpha^2F(\omega)}{\omega}d\omega},
\end{equation}
\begin{equation}
\alpha^2F(\omega)=\frac{1}{2{\pi}N(\varepsilon_F)}\sum_{{\bf q}\nu}\delta(\omega-\omega_{{\bf q}\nu})\frac{\gamma_{{\bf q}\nu}}{\hbar\omega_{{\bf q}\nu}},
\end{equation}
where N($E_F$) is the density of states at the Fermi level, $\omega_{{\bf q}\nu}$  is the frequency of the $\nu$-th phonon mode at the wave vector {\bf q}, and  $\gamma_{{\bf q}\nu}$ is the phonon linewidth~\cite{Eliashberg},
\begin{equation}
\gamma_{{\bf q}\nu}=2\pi\omega_{{\bf q}\nu}\sum_{{\bf k}nn'}|g_{{\bf k+q}n',{\bf k}n}^{{\bf q}\nu}|^2\delta(\varepsilon_{{\bf k}n}-\varepsilon_\text{F})\delta(\varepsilon_{{\bf k+q}n'}-\varepsilon_\text{F}),
\end{equation}
in which $g_{{\bf k+q}n',{\bf k}n}^{{\bf q}\nu}$ is an EPC matrix element. The logarithmic average frequency $\omega_{log}$  is defined as
\begin{equation}
\omega_{log}=\text{exp}[\frac{2}{\lambda}\int{\frac{d\omega}{\omega}\alpha^2F(\omega){ln}(\omega)}].
\end{equation}

\section{Results and Analysis}

\begin{figure}[t!]
\centering
\includegraphics[width=0.65\columnwidth]{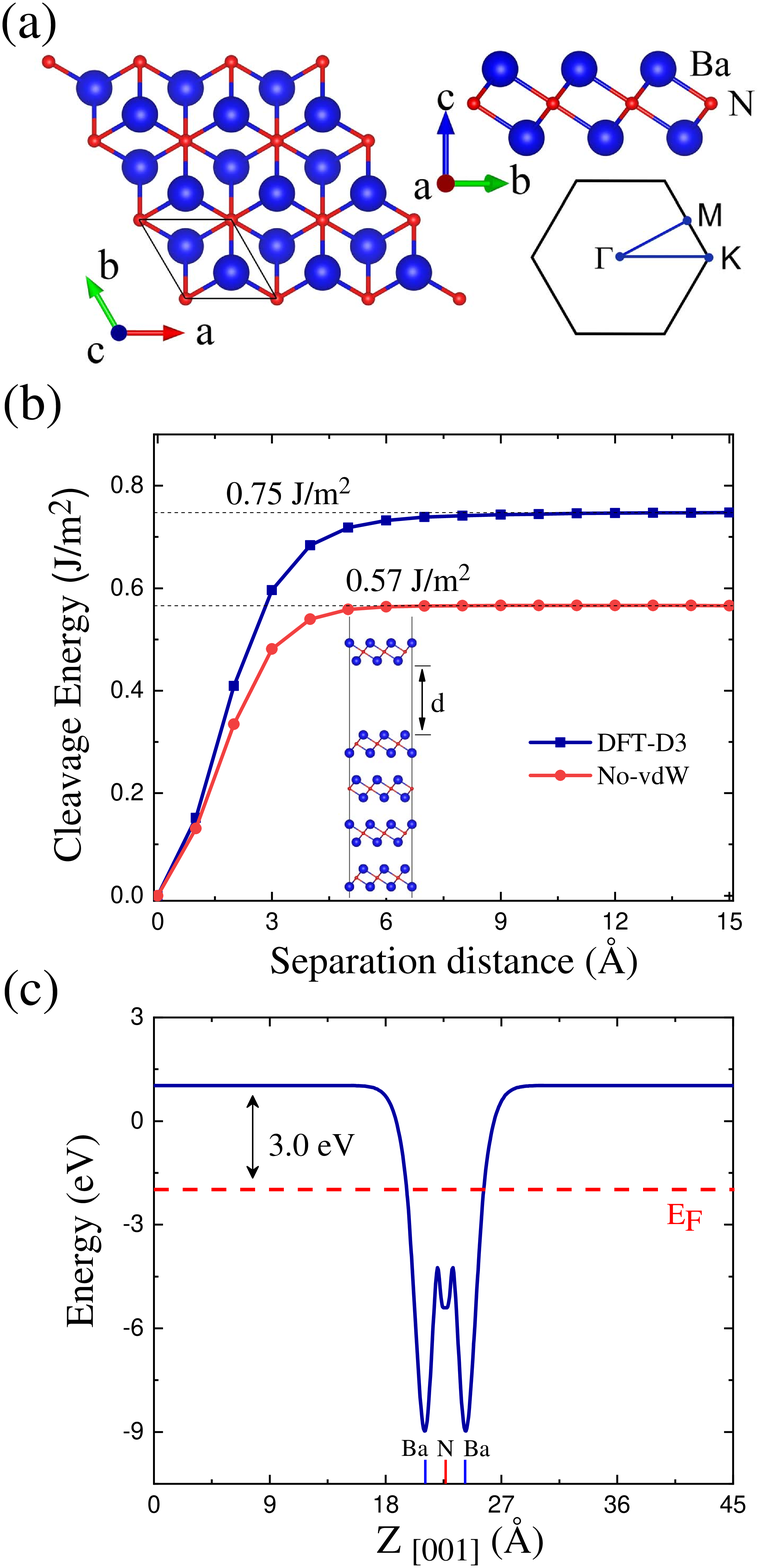}
\caption {(a) Top and side views of monolayer Ba$_2$N. The blue and red balls represent Ba and N atoms, respectively. The unit cell is marked by the black solid line. The lower right corner shows the two-dimensional (2D) Brillouin zone (BZ) of the unit cell. (b) Cleavage energy of Ba$_2$N calculated with and without the van der Waals correction. The inset shows the schematic picture for the cleavage process. (c) Work function of monolayer Ba$_2$N. The red dotted line labels the Fermi level. The atomic positions are marked by the color bars on the horizontal axis.}
\label{fig:1}
\end{figure}

The top and side views of the optimized crystal structure of monolayer Ba$_2$N are displayed in Fig. \ref{fig:1}(a). Apparently, monolayer Ba$_2$N has three atomic planes with the stacking sequence of Ba-N-Ba and its point group symmetry is D$_{3d}$. The calculated in-plane lattice constant of monolayer Ba$_2$N is 4.02 {\AA}, which is consistent with the experimental value~\cite{O. Reckeweg}. In order to examine whether or not bulk Ba$_2$N can be easily exfoliated into the monolayer form, we calculated the cleavage energy with and without the van der Waals (vdW) correction~\cite{DFT-D3}. As shown in Fig. 1(b), the cleavage energy calculated with (without) the vdW interaction is 0.75 (0.57) J/m$^{2}$, which is lower than that of Ca$_2$N (1.09 J/m$^{2}$)~\cite{S.-T. Zhao}. Therefore, Ba$_2$N may be cleaved into monolayer via the liquid exfoliation method as Ca$_2$N~\cite{D. L. Druffel}. The calculated work function of monolayer Ba$_2$N is 3.0 eV [Fig. \ref{fig:1}(c)], being comparable with the one of Y$_2$C (2.9 eV)~\cite{X. Zhang} and slightly higher than that of Ca$_2$N (2.6 eV)~\cite{K. Lee}. This reflects the characteristics of 2D electrides and suggests that Ba$_2$N can serve as a good electron donor as a substrate.

\begin{figure}[!t]
\centering
\includegraphics[width=1.0\columnwidth]{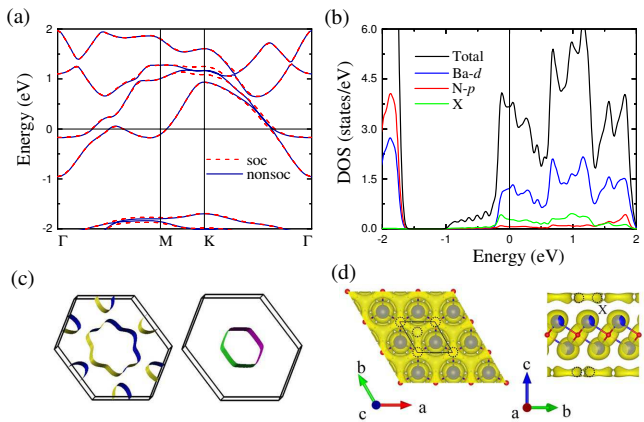}
\caption {(a) Band structure, (b) density of states (DOS), and (c) Fermi surface of monolayer Ba$_2$N. (d) Top and side views of the electron localization function (ELF) maps for monolayer Ba$_2$N with an isosurface value of 0.5. The dashed circles represent the positions of pseudoatoms.}
\label{fig:2}
\end{figure}

The band structure of monolayer Ba$_2$N is exhibited in Fig. \ref{fig:2}(a). There are two bands across the Fermi level, indicating the metallic behavior of monolayer Ba$_2$N. In consideration of the large atomic mass of Ba, we also performed the calculations with the spin-orbit coupling (SOC). As can be seen, the SOC has a minor effect on the bands around the Fermi level and we thus will not consider it in the following discussion. The corresponding Fermi surface of monolayer Ba$_2$N is shown in Fig. \ref{fig:2}(c). Clearly, there are electron-type pockets around both $\Gamma$ and M points, which is different from the case of Ca$_2$N in which there are Fermi pockets only at the $\Gamma$ point~\cite{S. Guan}. The top and side views of the electron localization function (ELF) map for monolayer Ba$_2$N are shown in Fig. \ref{fig:2}(d). There is free electron gas floating on the upper and lower surfaces, which is similar to monolayer Ca$_2$N~\cite{X.-L. Qiu}. To further study the electronic structure of monolayer Ba$_2$N, four pseudoatoms (labeled as X) with Wigner-Seitz radii of 1.1 {\AA} were added to the electron gas as labeled by the dashed circles in the ELF map [Fig. \ref{fig:2}(d)]. From the projected density of states (DOS) shown in Fig. \ref{fig:2}(b), we learn that the bands around the Fermi level are mainly contributed by the $5d$ orbitals of Ba atoms and the free electron gas. According to the ionic valence states of Ba$^{2+}$ and N$^{3-}$, the free electron gas may come from the 6$s$ orbitals of Ba atoms. Due to large distance of the free electron gas from the Ba planes [Fig. 2(d)], it cannot be well described by the atomic orbital basis. Moreover, since the pseudoatoms do not cover all of the free electron gas [Fig. 2(d)], the contribution of the electron gas to the DOS is much more than that displayed in Fig. 2(b).

\begin{figure}[!t]
\centering
\includegraphics[width=1.0\columnwidth]{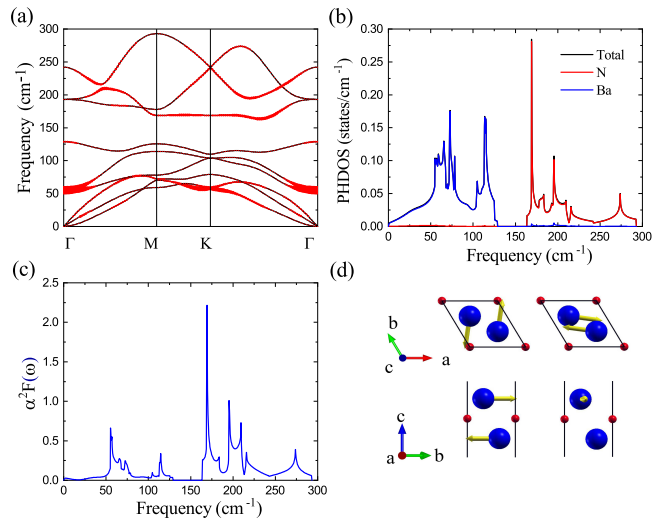}
\caption {(a) Phonon dispersion of monolayer Ba$_2$N. The sizes of red dots are proportional to the phonon linewidth $\gamma_{{\bf q}\nu}$. (b) Phonon density of states (PHDOS). (c) Eliashberg spectral function $\alpha^2F(\omega)$. (d) Vibrational patterns of the optical phonon modes with the frequency of 55 cm$^{-1}$ around the $\Gamma$ point. The length of the arrows indicates the amplitude of the atomic motion.}
\label{fig:3}
\end{figure}

Figure \ref{fig:3} shows the phonon dispersion and the EPC property of monolayer Ba$_2$N. From Fig. \ref{fig:3}(a), the absence of imaginary frequency in the phonon dispersion across the whole Brillouin zone indicates the dynamical stability of monolayer Ba$_2$N. Based on the calculated phonon linewidths $\gamma_{{\bf q}\nu}$, there are large contributions to the EPC strength coming from the optical phonon branches with the frequency of 55 cm$^{-1}$ around the $\Gamma$ point. The vibrational patterns of these phonon modes are demonstrated in Fig. \ref{fig:3}(d). Interestingly, both modes show opposite vibrations of Ba atoms in the $x$-$y$ plane. From the phonon density of states [Fig. \ref{fig:3}(b)], the phonons with frequencies below 130 cm$^{-1}$ mainly originate from the vibrations of Ba atoms. In comparison, the phonon modes in the frequency range of 160 to 220 cm$^{-1}$ are mainly contributed by the N vibrations [Fig. \ref{fig:3}(b)] and show sharp peaks in the Eliashberg spectral function $\alpha^2F(\omega)$  [Fig. \ref{fig:3}(c)]. Based on the McMillan-Allen-Dynes formula, the calculated superconducting $T_c$ of monolayer Ba$_2$N is 3.4 K. To double check the superconductivity of monolayer Ba$_2$N, we also evaluate the superconducting $T_c$ by using the anisotropic Migdal-Eliahberg theory~\cite{Giustino, Margine}, as shown in Fig. \ref{fig:7} in the Appendix. Clearly, the anisotropic superconducting gap vanishes around 6 K, which is higher than that calculated with the McMillan-Allen-Dynes formula (3.4 K). The similar situations also occur in other systems~\cite{You, Wu, Chen} and hence our calculation result on the superconductivity of monolayer Ba$_2$N is reasonable.

\begin{figure}[t]
\centering
\includegraphics[width=1.0\columnwidth]{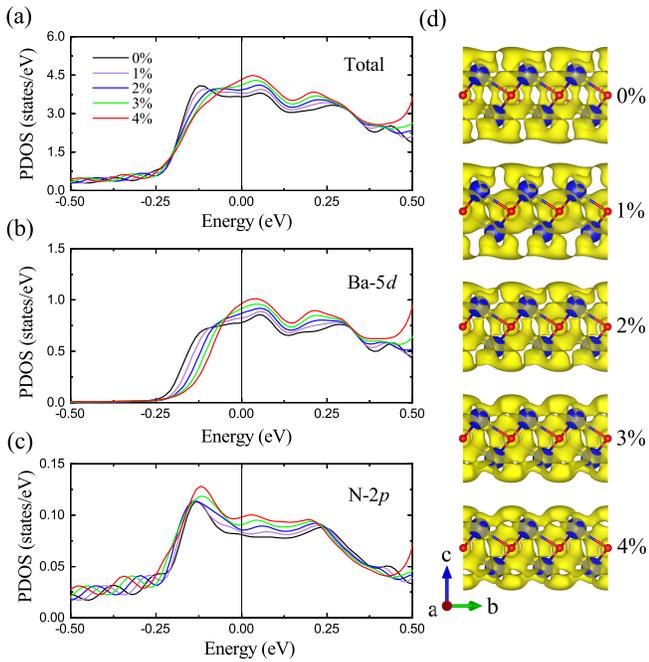}
\caption {(a) Total DOS, (b) Partial DOS from Ba-$5d$ orbitals, and (c) Partial DOS from N-$2p$ orbitals under different biaxial tensile strains. (d) Partial charge densities in the energy range of -0.1 to 0 eV at different strains with an isosurface value of 0.0005 e/{\AA}$^{3}$.}
\label{fig:4}
\end{figure}

We further studied the effect of tensile strain on the electronic structure of monolayer Ba$_2$N. The in-plane biaxial tensile strains within 4$\%$ have been considered, which are accessible experimentally. Figure \ref{fig:4}(a)-\ref{fig:4}(c) show the total DOS and partial DOSs from Ba-5$d$ and N-2$p$ orbitals at different strains, respectively. It is clear that the DOSs at the Fermi level increase gradually with the enlarged tensile strains, which facilitates the coupling of more electronic states with the phonons. The calculated partial charge densities around the Fermi level ([-0.1, 0] eV) in Fig. \ref{fig:4}(d) demonstrate that with the increasing tensile strain the free electron gas floating on the upper and lower surfaces of monolayer Ba$_2$N accumulates into the intra-layer region, which coincides with the increase of PDOSs from Ba and N atoms around the Fermi level.

\begin{figure}[h]
\centering
\includegraphics[width=0.67\columnwidth]{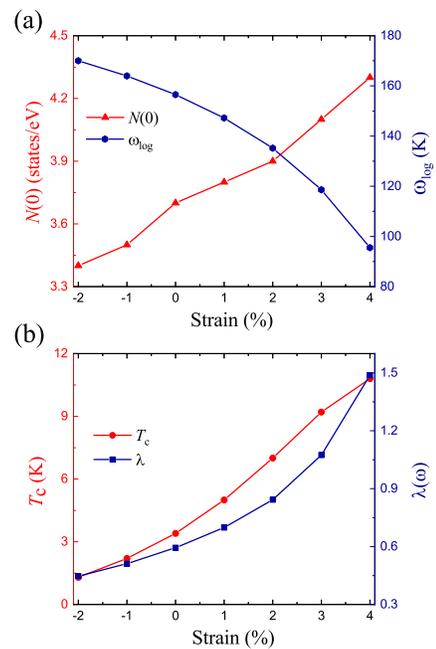}
\caption {(a) DOS at the Fermi level $N(0)$ (red) and the logarithmic average frequency $\omega_{log}$ (blue) of monolayer Ba$_2$N at different strains. (b) Calculated superconducting transition temperature $T_c$ (red) and EPC constant $\lambda$ (blue) of monolayer Ba$_2$N at different strains.}
\label{fig:5}
\end{figure}

The superconducting properties of monolayer Ba$_2$N also show dramatic changes with the lattice strain. In Fig. \ref{fig:5}, we show the effect of strains from -2$\%$ (compressive strain) to 4$\%$ (tensile strain) on the DOS at the Fermi level $N(0)$, the logarithmic average frequency $\omega_{log}$, the integrated EPC constant $\lambda$, and the superconducting transition temperature $T_c$. The compressive strain has a negative effect on the superconducting properties, because it reduces the DOS at the Fermi level and the EPC intensity. In comparison, the tensile strain brings an increase in the EPC constant from $\lambda$=0.59 in the strain-free case to $\lambda$=1.49 at a 4$\%$ tensile strain. According to the McMillan-Allen-Dynes formula, the superconducting temperature $T_c$ is related to both the EPC constant $\lambda$ and the logarithmic average frequency $\omega_{log}$. Although $\omega_{log}$ decreases with the tensile strain, the compensation effect from $\lambda$ is reflected in the behavior of the superconducting $T_c$, which rises from 3.4 K without the strain to 10.8 K at a 4$\%$ tensile strain. Further tensile strain beyond 5$\%$ will induce the dynamical instability of monolayer Ba$_2$N, as shown in Fig. \ref{fig:8} of the Appendix. We also evaluate the impact of empirical $\mu^*$   values, which have small influence on $T_c$ and do not change the relationship between $T_c$ and lattice strain (Fig. \ref{fig:9} in the Appendix). The detailed reason for the increasing $T_c$ with the moderate tensile strain will be explained below.

\begin{figure}[h]
\centering
\includegraphics[width=1.0\columnwidth]{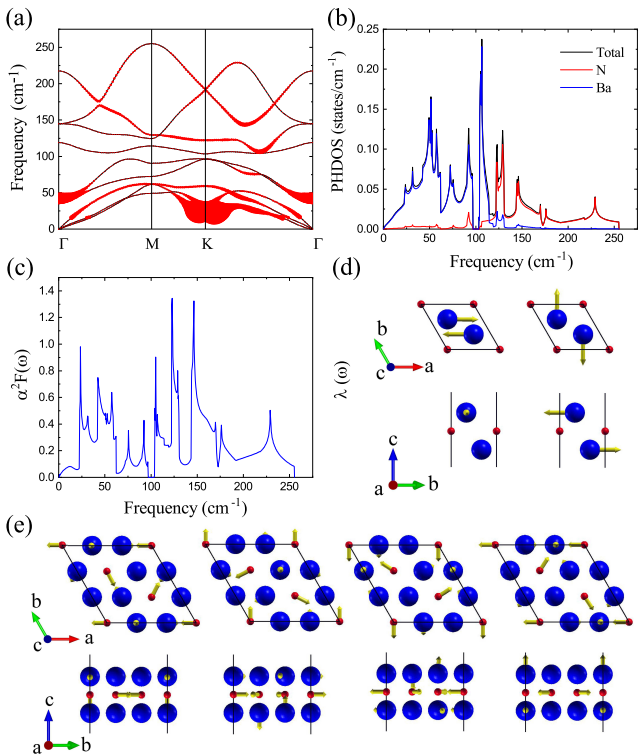}
\caption {(a) Phonon dispersion of monolayer Ba$_2$N at 4$\%$ tensile strain. The sizes of red dots are proportional to the phonon linewidth $\gamma_{{\bf q}\nu}$. (b) PHDOS. (c) Eliashberg spectral function $\alpha^2F(\omega)$. The vibration patterns of (d) the optical phonon modes with the frequency of 49 cm$^{-1}$ around the $\Gamma$ point and (e) the softened acoustic phonon modes around the frequency of 24 cm$^{-1}$ at the K point. The length of arrows indicates the amplitude of atomic motions.}
\label{fig:6}
\end{figure}

We choose the case of 4$\%$ tensile strain as an example to illustrate the origin underlying the increased $T_c$. The phonon dispersion of monolayer Ba$_2$N at 4$\%$ tensile strain is shown in Fig. \ref{fig:6}(a). Similar to the scene without strain, there is a strong EPC for the phonons with frequencies of 49 cm$^{-1}$ at the $\Gamma$ point, which involve the opposite vibrations of Ba atoms in the $x$-$y$ plane [Fig. \ref{fig:6}(d)]. Meanwhile, the phonon modes in the frequency range of 120 to 172 cm$^{-1}$ across the BZ, which mainly originate from the N vibrations [Fig. \ref{fig:6}(b)], also contribute significantly to the EPC and exhibit sharp peaks in the Eliashberg spectral function $\alpha^2F(\omega)$ [Fig. \ref{fig:6}(c)]. In addition to these similarities, there are some significant differences between the 4$\%$ tensile strain case and the strain-free one. As the tensile strain increases, the phonon modes generally shift towards lower frequencies with the significant softening of the acoustic branch at the K point [Fig. \ref{fig:6}(a)]. The softened modes at the K point correspond to the sharp peaks around 24 cm$^{-1}$ in the Eliashberg spectral function $\alpha^2F(\omega)$.
Besides the peak at 24 cm$^{-1}$ in Fig. \ref{fig:6}(c), there are two new small peaks at 72 and 92 cm$^{-1}$, which correspond to the hybridization peaks in the PHDOS of Ba and N atoms [Fig. \ref{fig:6}(b)]. Furthermore, the peak at 105 cm$^{-1}$ has a significant enhancement [Fig. \ref{fig:6}(c)], and there is also hybridization between the PHDOS of Ba and N atoms [Fig. \ref{fig:6}(b)].
In order to obtain the vibration patterns of the softened acoustic branch with the large EPC at the K point [Fig. \ref{fig:6}(a)], we expanded the unit cell to a $\sqrt{3}\times\sqrt{3}$ supercell, and then the K point of unit-cell BZ folds to the $\Gamma$ point of the supercell BZ. It is obvious that in this softened phonon mode the Ba atoms have both in-plane and out-of-plane vibrations, while the N atoms merely show in-plane vibrations [Fig. \ref{fig:6}(e)]. This generates the strong EPC and increases the superconducting $T_c$ in the tensile strain case.

\section{DISCUSSION AND CONCLUSION}
Since Ba2N shares the same sandwich-like structure with 1T-TMDs and 2-1 MXenes, it is interesting to make a comparison of their electronic and phonon properties. Our above calculations indicate that in Ba$_2$N the Ba-5$d$ orbitals and free electron gas contribute most to the energy bands around the Fermi level, and the tensile strain retrieves more electrons into the intralayer region and strengthens their coupling with Ba and N vibrations. As a result, the superconducting $T_c$ of Ba$_2$N can be much enhanced with the tensile strain. In comparison, for the typical TMD materials, 1T-TaSe$_2$ and 1T-NbSe$_2$ monolayers have a Mott insulating state with a Star-of-David ($\sqrt{13}\times\sqrt{13}$) charge-density-wave (CDW) lattice distortion~\cite{Y. Chen, M. Calandra}, while 1H-TaSe$_2$ and 1H-NbSe$_2$ monolayers exhibit superconductivity ($T_c$ = 2.2 K for 1H-TaSe$_2$ and 3 K for 1H-NbSe$_2$) and a $3\times3$ CDW order~\cite{X.-X. Xi, C.-S. Lian}. Previous calculations show that for the band structure Ta-5$d$ orbitals in 1H-TaSe$_2$ and Nb-4$d$ orbitals in 1H-NbSe$_2$ dominate around the Fermi level, while the inner Ta or Nb in-plane atomic vibrations contribute most to the total EPC~\cite{C.-S. Lian, C.-S. Lian2}. As for MXene, both experiment and theory suggest that it is an ideal kind of material for discovering superconductors. However, until now, very few nitride MXenes have been synthesized~\cite{P. Urbankowski, N. M. Abbasi}  and among them only Mo$_2$N was predicted to be superconducting with a $T_c$ of 16 K ~\cite{mo2n}. The phonon properties of Mo$_2$N and Ba$_2$N are quite different. In Mo$_2$N, the phonon frequencies from Mo and N atoms are almost continuously connected due to the large dispersion of the N-based phonon modes~\cite{mo2n}. In contrast, there is a clear gap between the Ba and N modes in the phonon dispersion of Ba$_2$N [Fig. \ref{fig:3}(b)]. With the tensile strain, the phonon gap in Ba$_2$N is reduced and the acoustic branches around the K point begin to soften [Fig. \ref{fig:6}(b)], resulting in the enhanced EPC and $T_c$. Therefore, we think that the existence of these differences gives Ba$_2$N unique characteristics that will make it an interesting platform for studying two-dimensional superconductors.

In summary, we have investigated the electronic and superconducting properties of monolayer electride Ba$_2$N by using first-principles electronic structure calculations. We find that the cleavage energy of bulk Ba$_2$N is 0.75 J/m$^{2}$, which is lower than that of bulk Ca$_2$N (1.09 J/m$^{2}$), giving the opportunity to cleave it into the monolayer by the liquid exfoliated method. The monolayer Ba$_2$N has a low work function of 3.0 eV and is predicted to be superconducting below 3.4 K. According to our analysis, the relative vibrations of Ba atoms on the upper and lower surfaces contribute most to the electron-phonon coupling. With the biaxial tensile strain, both the electronic DOS at the Fermi level and the EPC strength increase, which then enhances the superconductivity. Notably, at the 4$\%$ tensile strain, the superconducting $T_c$ rises to 10.8 K, where the in-plane vibrations of N atoms also play an important role in the EPC. The tunable superconductivity in monolayer electride Ba$_2$N thus provides a platform to explore more superconductors in the ultrathin limit.

\begin{figure}[!t]
\centering
\includegraphics[width=0.65\columnwidth]{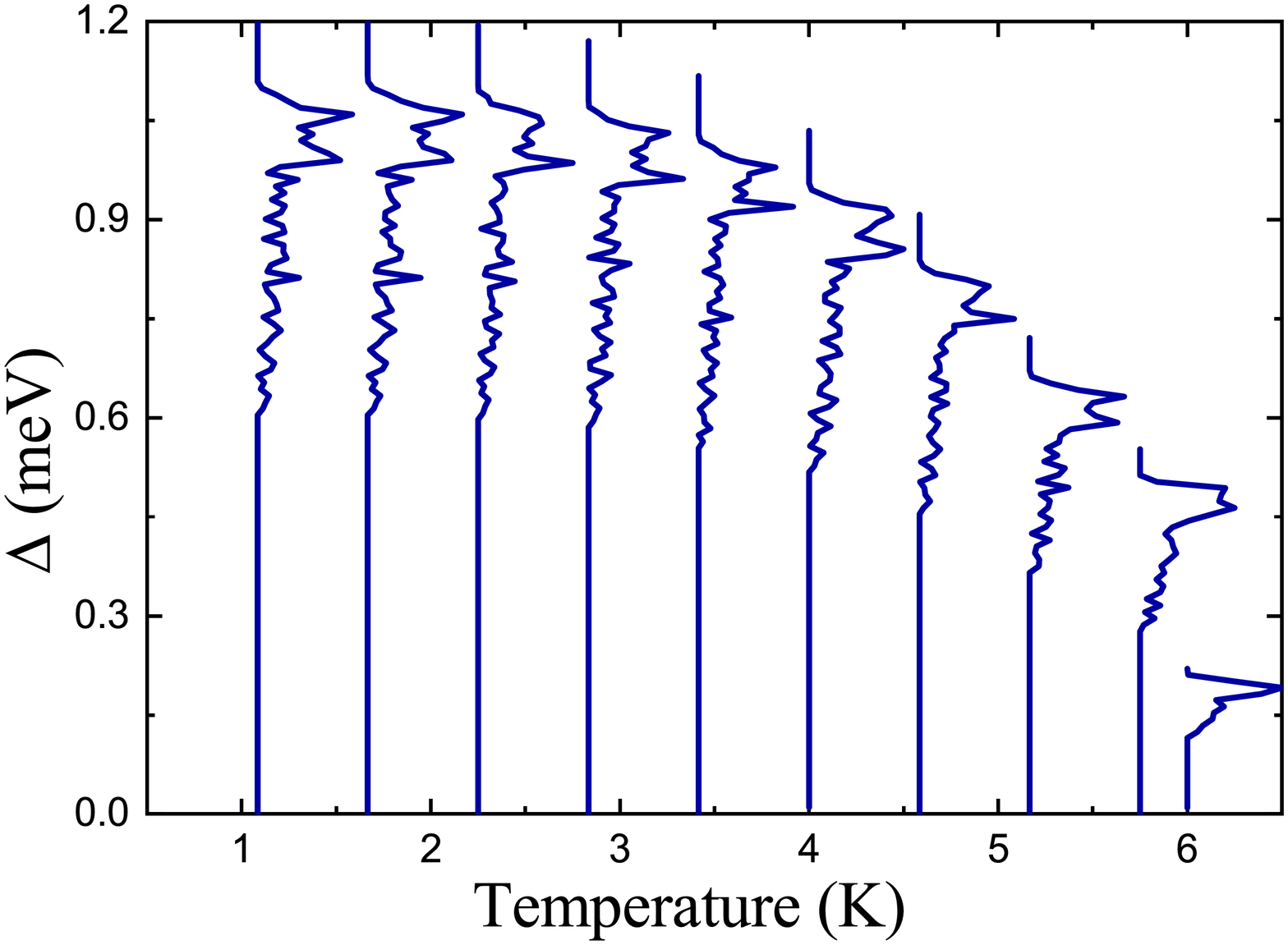}
\caption {Anisotropic superconducting gap of monolayer Ba$_2$N as a function of temperature.}
\label{fig:7}
\end{figure}

\begin{acknowledgments}
We wish to thank C. C. Liu for stimulating discussion. This work was supported by the National Key R\&D Program of China (Grants No. 2017YFA0302903 and No. 2019YFA0308603), the Beijing Natural Science Foundation (Grant No. Z200005), the National Natural Science Foundation of China (Grants No. 11774424, No. 11934020, and No. 12174443), the CAS Interdisciplinary Innovation Team, the Fundamental Research Funds for the Central Universities, and the Research Funds of Renmin University of China (Grants No. 19XNLG13 and No. 20XNH064). Computational resources were provided by the Physical Laboratory of High Performance Computing at Renmin University of China.
\end{acknowledgments}

\begin{appendix}
\section{}

\begin{figure}[thb]
\centering
\includegraphics[width=0.65\columnwidth]{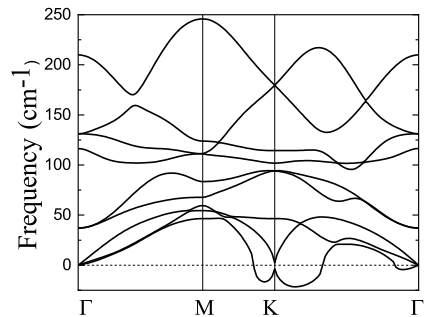}
\caption {Phonon dispersion of monolayer Ba$_2$N at 5$\%$ tensile strain.}
\label{fig:8}
\end{figure}

\begin{figure}[h]
\centering
\includegraphics[width=0.65\columnwidth]{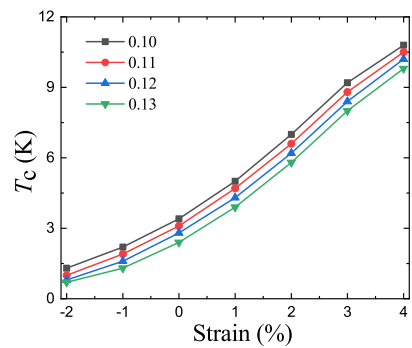}
\caption {Relationship between superconducting $T_c$ and applied tensile strain for monolayer Ba$_2$N with the $\mu^*$ values varying from 0.10 to 0.13.}
\label{fig:9}
\end{figure}

Figure \ref{fig:7} shows the temperature evolution of the anisotropic superconducting gap for strain-free Ba$_2$N monolayer, which was calculated with the anisotropic Migdal-Eliahberg theory as implemented the EPW code ~\cite{Giustino, Margine}. 

At 5\% tensile strain, the phonon dispersion of monolayer Ba$_2$N shows imaginary frequencies around the K point (Fig. \ref{fig:8}), indicating the dynamical instability.

Figure \ref{fig:9} displays the dependence of superconducting $T_c$ on the applied tensile strain calculated with the $\mu^*$ values varying from 0.10 to 0.13. It can be seen that the adopted $\mu^*$ values have little influence on the tendency of $T_c$ and do not change the conclusion that the tensile strain can significantly increase $T_c$.

\end{appendix}

\end{document}